\documentclass[aps,twocolumn,prd,showpacs,nofootinbib]{revtex4}
\usepackage{amsmath}
\usepackage{graphicx}
\usepackage{dcolumn}
\usepackage{bm}
\usepackage{amssymb}
\usepackage{latexsym}

\def\be{\begin{equation}}
\def\ee{\end{equation}}
\def\ba{\begin{eqnarray}}
\def\ea{\end{eqnarray}}

\begin{document}

\title{Nearly Divergence of Correlation Length and Perturbation Spectrum in
String Gas Cosmology }

\author{Yun-Song Piao}
\affiliation{} \affiliation{College of Physical Sciences, Graduate
School of Chinese Academy of Sciences, YuQuan Road 19A, Beijing
100049, China}

\begin{abstract}

Recently, it has been shown in Ref. \cite{NBV} that the string
thermodynamic fluctuation may lead to a scale invariant spectrum
of scalar metric perturbation. However, its realization is still
in study. In this note we suppose that the correlation length of
metric perturbation, which is proportional to the sound speed,
might be nearly divergent at the critical point of phase
transition. In this case we find that the string gas mechanism
responsible for the generation of
primordial perturbation may be applied well. 

\end{abstract}


\maketitle

In string gas cosmology \cite{BV, TV}, see Refs. \cite{BW} and
\cite{B} for recent reviews, it is assumed that the universe
starts in a Hagedorn phase, in which the universe is initially in
thermal equilibrium at a temperature close to the Hagedorn
temperature, the limiting temperature of perturbative string
theory. In this case, by using the tools of string thermodynamics
it was showed that the nearly scale invariant spectrum of
cosmological fluctuations responsible for the structure formation
of observable universe may be obtained \cite{NBV}, see Ref.
\cite{BNPV} for the tensor perturbation and also Refs. \cite{N,
BNPV2, BKSE} for more details.

However, to make this seeding mechanism feasible, it seems that
the dilaton need to be fixed, which may be argued with the strong
coupling Hagedorn phase \cite{BKSE}. If the dilaton is not fixed,
the terms related to the dilaton running will dominate the (00)
equation of metric perturbation $\Phi$. In this case instead scale
invariant spectrum one will obtain a Poisson spectrum \cite{BKSE},
When the dilaton is fixed, the perturbation equations are those of
Einstein gravity. In this case, the perturbation equation
$k^2\Phi\sim \delta\rho$ may be used only when $k>h$, in which $k$
and $h$ are the comoving wave number and comoving Hubble parameter
in the Einstein frame respectively. However, in Ref. \cite{NBV},
for fixed dilaton, it seems that the scales of metric perturbation
interested today are generally super Hubble radius during Hagedorn
phase, i.e. $k<h$ \cite{KKLM}. The metric perturbation can be
driven by the matter fluctuation only on scale smaller than the
Hubble radius, while on super Hubble scale the metric perturbation
is dominated since the matter oscillations have frozen out. Thus
one is not able to compute the matter fluctuations on such scales
and to then use them to induce the metric fluctuations.
Thus in this case the calculations of metric perturbation
seems be controversial 
\cite{KKLM}. However, one may relax these controversies by
assuming that the strong Hagedorn phase lasts sufficiently long
\cite{BKSE} or invoking a bounce cosmology \cite{BBMS} and also
\cite{AKK}.

In this note, we propose a different possibility. We suppose that
the correlation length $l\sim c_sh^{-1} $ of metric perturbation
might be nearly diverged at the critical point of phase
transition, where $c_s$ denotes the sound speed of metric
perturbation. When ${\cal T}\rightarrow {\cal T}_c$, where ${\cal
T}_c$ is the critical temperature of phase transition, since the
change of $h$ is generally expected to be moderate during phase
transition \footnote{see Refs. \cite{BBMS, BMS} for a different
case in which the bounce is introduced during phase transition,
which actually also corresponds to one of the cases with diverged
correlation length since $h= 0$ at some epoch of the bounce. },
the divergence of correlation length of metric perturbation
suggests that $c_s({\cal T})$ will rapidly increase to infinity
with ${\cal T}$, see Fig.1. Based this supposition we find those
controversial issues relevant to the string gas mechanism
responsible for the generation of primordial perturbation can be
solved well. For some ansatz of correlation length, we calculate
the spectrum of scalar metric perturbation by using the tools of
string thermodynamics, as in Ref. \cite{NBV}. The spectral index
generally has a moderate red tilt, which may be consistent with
recent observations well. For an illustration we firstly assume
that the background dilaton velocity is negligible. We will relax
this assumption later. Besides, here we will not involve relevant
contents with the stability of moduli, which may be seen in Refs.
\cite{PB, BCW} in the string gas cosmology.

When the dilaton is fixed, the background and perturbation
equations
are those of Einstein gravity. 
Though the nearly divergence of sound speed can hardly be
understood in Einstein gravity and is generally expected to have a
profound origin, for a phenomenological discussion we may
implemented it by e.g. endowing the Einstein action with an added
term \be \sim\int d^4x\sqrt{-g}c_s^{2}{\cal R}^{(3)}
,\label{add}\ee where ${\cal R}^{(3)}_{ij}$ is the Ricci curvature
described by the space component $g_{ij}$ of metric $g_{\mu\nu}$
and its corresponding affine connections. In this sense ${\cal
R}^{(3)}$ is actually the intrinsic spatial curvature on constant
time slicing. The factor $c_s^2$ only depends on the temperature,
when ${\cal T}\sim {\cal T}_c$, it is required to be very large,
while ${\cal T}\ll {\cal T}_c$, we expect that the Einstein
gravity should be resumed, thus it should approach $0$. It seems
that this term breaks the Lorentz invariance, however, it can be
recovered after the transition, i.e. ${\cal T}\ll {\cal T}_c$,
which is consistent with present observations. By making the
variation of (\ref{add}), we have \ba \delta (\sqrt{-g}{\cal
R}^{(3)}) & \rightarrow & ({1\over 2}\sqrt{-g}g_{\mu\nu}{\cal
R}^{(3)})\delta g^{\mu\nu}+ \nonumber\\
&  & \sqrt{-g}\delta(g^{ij}{\cal R}^{(3)}_{ij}), \label{delta}\ea
and can find that the (00) component of Einstein perturbation
equation will have an added correction
$\sim c_s^2 g_{\mu\nu}{\cal R}^{(3)}$
leaded by the first term of Eq.(\ref{delta}). Though there is also
a term $\sim\delta (g^{ij}{\cal R}^{(3)}_{ij})$, it dose not
contribute the (00) component. The interest of this extra
correction lies in that in the longitudinal gauge, see Ref.
\cite{MFB} for details on the theory of cosmological
perturbations, it brings an added term $\sim c_s^2\nabla^2\Phi$ to
the $(00)$ equation of the metric perturbation $\Phi$, while does
not modify the background equation mastering the evolution of
scale factor $a$. This is consistent with the feature that ${\cal
R}^{(3)}$ is the intrinsic spatial curvature on constant time
slicing. However, it should be noted that to depict the nearly
divergence of correlation length during the transition we
introduce the term (\ref{add}), which, however, is not necessary
and exclusive, there may also be other possible ways to do so.
Here we only take such an example to show the feasibility of our
supposition.

When $c_s$ is very large,
i.e. $c_sk\gg h$, which means that the perturbations are very deep
into the sound horizon and thus the terms $h^2\Phi$ and
$h\Phi^{\prime}$ may be neglected, in the longitudinal gauge the
$(00)$ equation of the metric perturbation can be reduced to \be
c_s^2k^2\Phi \simeq a^2G\delta\rho, \label{c2k2}\ee which is
Poisson like, but is in relativistic sense. We can obtain
Eq.(\ref{c2k2}) completely contributes to the introduction of
added term to Einstein action, since it brings a nearly diverged
sound speed when the temperature approaches the critical point.
Thus we have \be {\cal P}_{\Phi}(k)\simeq {a^2G^2\over c_s^4
k^4}{\cal P}_{\delta \rho}(k)={a^2G^2\over c_s^4 k^4}<\delta
\rho^2>_{R=a/(c_sk)}, \label{pk0}\ee where $R$ is the size of
region in which the
fluctuations are calculated. 
In Eq.(\ref{pk0}), it is obviously seen that we replace $R=a/k$
used in Ref. \cite{NBV} with $R=a/(c_sk)$. The reason is in the
following. What we suppose here is only the nearly divergence of
correlation length of metric perturbation. That of matter
fluctuation is still limited by Hubble scale during phase
transition. Thus in principle one need to calculate the
fluctuations of the energy momentum tensor of stringy matter on
various length $R$, up to the physical Hubble scale $a/h$.
However, here though the physical wavelength $a/k$ of matter
fluctuations required to induce the metric perturbation at present
observable scale are sub sound horizon scale, they are generally
super Hubble scale, i.e. $k<h$, see `A' mode denoted by the red
solid lines in Fig.1, thus in this case we can not deduce the
metric perturbation by calculating the matter fluctuation, since
in super Hubble scale the oscillating of matter fluctuations are
freezed. However, after we introduce the nearly divergent sound
speed, the effective physical wavelength of metric perturbation
will obtain a strong suppression $\sim 1/c_s$ and become $ a/(c_s
k)$, which may be sub Hubble scale well, see `A$^\prime$' mode
denoted by the red dashed lines in Fig.1. Thus in this case we may
take $R=a/ (c_s k)$ as length scale to calculate the matter
fluctuations and then use them to deduce the metric perturbations
by Eq.(\ref{pk0}), up to $a/h$.

This can also be explained in another perspective. We may define
the effective comoving wave number as $\omega=c_sk$, which is
rapidly decreased during phase transition due to the change of
$c_s$. Thus the evolution of comoving wavelength $\sim 1/c_s$ of
corresponding mode can be much faster than that of the Hubble
radius, which will make relevant mode be able to leaving the
horizon during phase transition, see `A$^\prime$' mode in Fig.1.
In this case initially the modes with the effective comoving
wavelength $1/\omega$ are well in the Hubble horizon, thus we may
calculate the matter fluctuation with length scale
$R=a/\omega=a/(c_sk)$ and then deduce the metric perturbation by
Eq.(\ref{pk0}). This explanation corresponds to that in the
perspective of the effective comoving wave number. The similar
analysis has appeared in Ref. \cite{Piaocs}. While in the
perspective of the correlation length or sound horizon in last
paragraph, the comoving wavelength of `A' mode in Fig.1 is
constant during phase transition, with the decreasing of sound
speed it will leave the sound horizon and become the primordial
perturbation. Note that the time when `A' mode leaves the sound
horizon, i.e. $k=h/c_s$, corresponds to that when `A$^\prime$'
mode leaves the Hubble horizon, i.e. $c_sk=h$, thus in this sense
both perspectives are actually equivalent, which can be seen in
Fig.1.


\begin{figure}[t]
\begin{center}
\includegraphics[width=7cm]{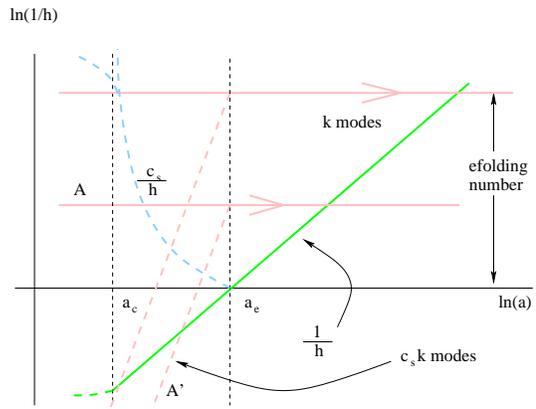}
\caption{The sketch of evolution of the sound horizon
$\ln{(c_s/h)}$ and the Hubble horizon $\ln{(1/h)}$ with respect to
the scale factor $\ln{a}$ during the transition. The transition in
which the winding modes of strings decay into radiation goes with
the rapid decreasing of sound speed, which makes the perturbations
be able to leave the sound horizon and become nearly scale
invariant primordial perturbations responsible for the structure
formation of observable universe. }
\end{center}
\end{figure}

During radiation domination, the energy is in the radiative
degrees of freedom, which correspond to the momentum modes of
strings. But when the temperature is about ${\cal T}_c$, it may be
expected that the oscillatory and winding modes of strings can be
excited, which will contribute most of energy in the string gas.
Thus in principle one is able to compute the spectrum of matter
fluctuation by using the result of closed string thermodynamics,
which is required to source the metric perturbation by
Eq.(\ref{c2k2}). Following \cite{NBV}, the perturbation of energy
density of closed string modes in the thermal equilibrium inside
an arbitrary volume $\sim R^3$ is given by \be
<\delta\rho^2>={{\cal T}^2\over R^6} c_{\rm V}, \label{drho}\ee
where the specific heat \be c_{\rm V}\simeq {R^2 {\cal T}_c^4\over
{\cal T}({\cal T}_c -{\cal T})},\ee which scales as $R^2$.
Eq.(\ref{drho}) indicates that the fluctuation of stringy matter
is a Poisson spectrum $\sim k^4$. This result is a key feature of
string thermodynamics, which was derived in Ref. \cite{DJNT} and
holds in the case of three large dimensions with the topology of a
torus. We substitute Eq.(\ref{drho}) into Eq.(\ref{pk0}), and then
can obtain \be {\cal P}_{\Phi}\simeq {{\cal T}_c^4\over
m_p^4}\left({{\cal T} \over {\cal T}_c -{\cal T}}\right),
\label{phi}\ee where $G=m_p$ has been used and the numerical
factor ${\cal O}(1-100)$, which depends on the rescale of Planck
scale $m_p$, has been neglected. Thus one can see that the Poisson
spectrum of string gas induces a scale invariant spectrum of
metric perturbation $\Phi$ by Eq.(\ref{pk0}), which is the same as
the result obtained in Ref. \cite{NBV}. The difference is that
here the metric perturbations calculated are in super Hubble
scale, i.e. $k<h$, but sub sound horizon scale, i.e. $k>h/c_s$.
During the transition in which the winding modes of strings decay
into radiation, we expect that $c_s$ decreases rapidly, as is
illustrated in Fig.1. In this case the metric perturbation can
exit the sound horizon, and after its leaving, it will quickly
freeze, since when $c_sk<h$, which means that the effective
wavelength of metric perturbation in perspective of effective
comoving wave number is larger than the horizon, the oscillating
of matter fluctuations has frozen. The spectrum of curvature
perturbation $\xi$ in comoving supersurface ${\cal P}_{\xi}\simeq
{\cal P}_{\Phi}$ up to a factor with order one, which is constant
in the super horizon scale. Thus the spectrum of the comoving
curvature perturbation can be nearly scale invariant and its
amplitude can be calculated at the time when the perturbation
exits the sound horizon, i.e. $k=h/c_s$, which gives the value of
${\cal T}$ in Eq.(\ref{phi}) at
the sound horizon crossing. 


The phase transition is not generally instantaneous, which will
lead to a tilt in the spectrum. This tilt depends on the change of
sound speed, since the change of sound speed determines the
evolution of sound horizon and so the sound horizon crossing time
of perturbation with some given wavelength $\sim k^{-1}$. In
principle to obtain the tilt of spectrum we need to know the
detailed evolution of sound speed with the temperature. However,
since we lack for the detailed knowledge of phase transition, here
we will appeal to some interesting ansatzs of $c_s$, which in some
sense might also help to our understanding for phase transition.
We firstly take \be c_s=\left({{\cal T}_c\over {\cal T}_c -{\cal
T}}\right)^p \label{cs1}\ee
as an attempt, where $p>0$ is a constant, and $p\rightarrow
\infty$ corresponds to an instantaneous transition. This ansatz in
some sense may be analogous to the case in condensed matter
physics, in which when the temperature approaches the critical
point of phase transition the correlation length of order
parameter diverges. For Eq.(\ref{cs1}), when ${\cal T}\rightarrow
{\cal T}_c$, the sound
speed diverges, 
while ${\cal T}\sim {\cal T}_e\ll {\cal T}_c$, where `e' denotes
the value at the end of transition, we have $c_s=1$, which means
the end of transition. In this time the perturbation equation is
that with the Einstein gravity.
The sound horizon crossing requires $k=h/c_s$, thus we can obtain
\be {k\over k_e} \simeq \left({{\cal T}_c\over {\cal T}_c - {\cal
T}}\right)^{-p},\label{kke1}\ee where we have neglected the change
of $h$ during phase transition, since the change of $h$ is much
smaller than that of $c_s$. When we include the change of $h$,
where we take $a\sim t^n$ and $n\sim {\cal O}(1/2)$ is a constant,
there will be a factor $({\cal T}_e/ {\cal T})^{(n-1)/n}$ before
the right hand term of Eq.(\ref{kke1}),which is negligible when
being compared to that of $c_s$.
We define \be {\cal N}\equiv\ln{\left({k_e\over k}\right)},
\label{N}\ee which measures the efolding number of mode with some
scale $\sim k^{-1}$ which leaves the sound horizon before the end
of transition, and thus $k_e$ means the last mode to be generated,
see Fig.1. When taking the comoving Hubble parameter $h=h_0$,
where the subscript `0' denotes the present time, we generally
have ${\cal N}\sim 50$, which is required by observable cosmology.
From Eq.(\ref{kke1}), we can see that when ${\cal T}\rightarrow
{\cal T}_c$, $k_e/k$ nearly approaches to infinity, thus the
efolding number is actually always enough as long as the
initial volume of early universe is large enough. 
By using Eq.(\ref{kke1}), the spectrum of metric perturbation can
be rewritten as \be {\cal P}_{\Phi} \simeq {{\cal T}_c^4\over
m_p^4}  \left({k\over k_e}\right)^{-{1\over
p}}\left[1-\left({k\over k_e}\right)^{1\over p}\right]^{-1}.
\label{phi1}\ee
Thus the amplitude is approximately ${\cal T}_c^4/m_p^4$ and the
spectral index is given by \be n_s -1= -{1\over
p}\cdot\left[{1\over 1-e^{-{\cal N}/ p}}\right],
\label{ns1}\ee 
where Eq.(\ref{N}) has been used. From Eq.(\ref{ns1}), we can see
that for fixed efolding number the spectral index is only
determined by the critical exponent $p$. To obtain the red tilt
required by the observations \cite{WMAP}, i.e. $n_s-1\simeq
-0.05$, for ${\cal N}\simeq 50$, it seems that we need $p\simeq
20$. In fact we also calculate the factor which is relevant to the
change of $h$ and has been neglected in Eq.(\ref{kke1}), and find
that it only contributes a term $\sim 1/p^2$, which is second
effect. Thus the approximation done in Eq.(\ref{kke1}) is
consistent.


In the ansatz (\ref{cs1}), the sound speed is diverged at the
critical point. It is also interesting to consider the case in
which the sound speed at the critical point arrives at a very
large value but not diverges. The example is given by \be c_s=
\left({{\cal T}\over {\cal T}_e}\right)^p ,\label{cs2}\ee where
when ${\cal T}\rightarrow {\cal T}_c$, $c_s$ approaches a
very large constant $({\cal T}_c/{\cal T}_e)^p$ since $p\gg 1$, 
while ${\cal T} \simeq {\cal T}_e$, we have $c_s\simeq 1$, which
means the end of transition. During the sound horizon crossing,
i.e. when $k=h/c_s$, we have \be {k\over k_e}\simeq \left({{\cal
T}_e\over {\cal T}}\right)^p,\label{kke2}\ee where similarly the
change of $h$ has been neglected. 
From Eq.(\ref{N}), we have $
{\cal N}=p\ln{({\cal T}/{\cal T}_e)}$. Note that though ${\cal
T}_e$ is generally not much less than ${\cal T}_c$, the enough
efolding number can still be assured, since $p\gg 1$. From
Eq.(\ref{phi}), the spectrum of metric perturbation is given by
\be {\cal P}_{\Phi} \simeq {{\cal T}_c^4\over m_p^4}
\left[\left({k\over k_e}\right)^{1\over p}\left({{\cal
T}_c\over {\cal T}_e}\right)-1\right]^{-1}. \label{phi2}\ee 
The spectral index is given by \ba n_s-1 & = & 
-{1\over p}\cdot\left({{\cal T}_c\over {\cal T}_c- {\cal
T}}\right)\nonumber\\ &=& -{1\over p}\cdot \left[{1\over 1-
e^{-({{\cal N}_c -{\cal N}\over p})}}\right], \label{ns2}\ea where
${\cal N}_c = \ln{(k_e/k_c)}$ is the total efolding number during
phase transition. Note that in Eq.(\ref{ns2}), when ${\cal
T}\rightarrow {\cal T}_c$, i.e. the total efolding number ${\cal
N}_c\simeq 50$, $n_s$ will be divergently red. Thus to match
Eq.(\ref{ns2}) to the observations, it seems that
$p\rightarrow \infty$ is required. However, 
if ${\cal N}_c$ is much larger than 50, even if $p$ does not
approach infinity, we also may have moderate red tilt of $n_s$ at
${\cal N}\simeq 50$.
For example, to obtain $n_s-1 \simeq -0.95$ required by the
observations, if we take $p\simeq 20$, the total efolding number
${\cal N}_c$ should be larger than at least $90$, which can
be seen in Eq.(\ref{ns2}). 


Now let us check whether the results are changed if the dilaton is
not fixed. When the dilaton is running, the background and
perturbation equations are those of dilaton gravity \cite{WB}.
Thus in principle we need to performed our calculations in the
string frame. However, note that the results of physical
quantities should be equivalent in the Einstein and string frames,
independent of the choice of frames. This suggests that all
calculations can be equally well implemented in both frames. Thus
for convenience we will continue our discussions in the Einstein
frame. The string frame metric can be related to the Einstein
frame metric by a conformal transformation ${\tilde
g}_{\mu\nu}=e^{2\varphi} g_{\mu\nu}$, where the tilde denotes the
quantity in the string frame and $\varphi$ is the dilaton. In the
string frame the thermal gas of string is coupled minimally to
gravity, thus after the conformal transformation, it will be
coupled to dilaton, and so can also serve for a source for the
dilaton perturbation while source for the metric perturbation. We
conformally introduce the added term $\sqrt{-{\tilde
g}}c_s^{2}{\cal {\tilde R}}^{(3)}$ in the action of string frame,
which, after the conformal transformation, leads to not only
$c_s^2 {\cal R}^{(3)}$ but $c_s^2(\nabla\varphi)^2$ in the
Einstein frame. Thus we see that not only metric perturbation may
be causally produced, but the dilaton perturbation.

In the (00) equation of metric perturbation, compared with the
case in which the dilaton is fixed, there will be some new terms,
such as $\varphi^{\prime 2}\Phi$,
$\varphi^{\prime}\delta\varphi^{\prime}$, which arise from the
running of dilaton. When these terms are dominated, instead
Eq.(\ref{c2k2}) one will approximately have $\Phi \sim
\delta\rho$, thus the Poisson spectrum of $\delta\rho$ will induce
a Poisson spectrum of $\Phi$, i.e. ${\cal P}_{\Phi}\sim k^4$, as
has been pointed out in Ref. \cite{BKSE}. Let us reexamine these
terms in our case. The term $\varphi^{\prime 2}\Phi$ can be
rewritten as $h^2(\Delta \varphi)^2 \Phi$, which reflects the
change of dilaton in unit of
Hubble time. 
Following \cite{BKSE}, when $\tilde t$ is large, one have $a\sim
{\tilde t}$ 
and also $e^{-\varphi}\sim {\tilde t}$, thus can obtain
$e^{-\varphi}\sim a$, which can be reduced to $\varphi \sim \ln{a}
$. Thus we can see that the $h^2(\Delta\varphi)^2$ is much smaller
than $c_s^2$ given by Eqs.(\ref{cs1}) or (\ref{cs2}). The similar
analysis also applies to $\varphi^{\prime}\delta\varphi^{\prime}$,
since it can be rewritten as $h^2\Delta\varphi\Delta\delta
\varphi$. These results indicate that compared with
$c_s^2k^2\Phi$, the terms related to the dilaton running are
negligible. Thus we can see that even if the dilaton is running,
Eq.(\ref{c2k2}) will be still approximately valid equation for
metric perturbation. Therefore we will still have a nearly scale
invariant spectrum (\ref{phi}) leaded by the Poisson like spectrum
(\ref{drho}) of string gas, whose tilt is given by Eq.(\ref{ns1})
or (\ref{ns2}).

The perturbation equation of dilaton in the Einstein frame has
also term $\sim c_s^2k^2 \delta\varphi_k$, which is due to the
appearance of $\sim c_s^2(\nabla\varphi)^2$ term after the
conformal transformation. This makes the causal generation of
dilaton perturbation become possible. In the perturbation equation
of dilaton, as long as the change of $\Phi$ in unite of Hubble
time is not far larger than $\delta\varphi_k $, we will have
$c_s^2k^2 \delta\varphi_k\gg h^2\Delta \varphi\Delta\Phi\simeq
\varphi^{\prime}\Phi^{\prime} $, which suggests that compared with
$c_s^2k^2\delta\varphi_k$, $\varphi^{\prime}\Phi^{\prime}$ term is
negligible. The other terms can also be neglected in the similar
analysis. Thus the perturbation equation of dilaton approximately
become $c_s^2k^2\varphi_k\sim \delta\rho$. This indicates that the
spectrum of dilaton perturbation is also scale invariant.

The Eq.(\ref{drho}) deduced by using the string thermodynamics
assumed the existence of thermal equilibrium in the Hagedorn
phase. However, as has been shown in Ref. \cite{DFM}, see also
\cite{EGJK, TK}, it seems that no such equilibrium can be builded
when the dilaton is running. The interaction rate of string
winding modes with the running of dilaton is generally more
rapidly changed than the physical Hubble parameter \cite{DFM}.
This means that when the dilaton is running, the gas of winding
strings will quickly fall out of the equilibrium. However, here
due to the introduction of the term $\sim c_s^2$, the causal
structure of spacetime has actually been changed. Thus in this
case it seems be required to rephrase the thermal equilibrium
condition. This is beyond the scope of this note, however, which
is worthy of further study.


In conclusion, 
we show that when the correlation length of metric perturbation,
which is proportional to the sound speed, at the critical point of
phase transition is nearly divergence, the string gas mechanism of
the generation of primordial perturbation
may be applied well. 
Though here we are constrained to the case with the added term
(\ref{add}), our work may be more general. The only requirement is
that the correlation length of metric perturbation is nearly
diverged at the critical temperature, which is significant to
obtain a causal structure responsible for the generation of
primordial perturbation and a Poisson like equation of metric
perturbation. In principle, this work may be applied to any cases
in which the Poisson like matter or energy fluctuation, like
Eq.(\ref{drho}), is required to induce the scale invariant
spectrum of metric perturbation by a Poisson like equation of
metric perturbation, like Eq.(\ref{c2k2}).
Finally, it should be pointed out that the phase transition that
the string winding modes become dominated is still a subject in
development and only partially understood at present. Thus this
study seems be slightly speculative, however, it might be helpful
for understanding the generation of primordial perturbation based
on the string thermodynamics and also the
physics of phase transition. 

\textbf{Acknowledgments} The author is grateful to Robert
Brandenberger for the insightful comments of earlier drafts and
discussions, and also Biao Jin for discussion on condensed matter
physics. This work is supported in part by NNSFC under Grant No:
10405029, in part by the Scientific Research Fund of
GUCAS(NO.055101BM03), as well as in part by CAS under Grant No:
KJCX3-SYW-N2.

\end{document}